\documentclass[preprint,showpacs,preprintnumbers,amsmath,amssymb]{revtex4}

\usepackage{graphicx}% Include figure files
\usepackage{epsfig}		
\usepackage{dcolumn}% Align table columns on decimal point
\usepackage{bm}% bold math

\def\0{\mbox{\tiny $0$}}
\def\1{\mbox{\tiny $1$}}
\def\2{\mbox{\tiny $2$}}
\def\3{\mbox{\tiny $3$}}
\def\4{\mbox{\tiny $4$}}
\def\5{\mbox{\tiny $5$}}
\def\6{\mbox{\tiny $6$}}
\def\7{\mbox{\tiny $7$}}
\def\8{\mbox{\tiny $8$}}
\def\9{\mbox{\tiny $9$}}

\def\n{\mbox{\tiny $n$}}
\def\k{\mbox{\tiny $k$}}

\def\f14{\mbox{\tiny $\frac{1}{4}$}}

\def\L{\mbox{\tiny $L$}}
\def\R{\mbox{\tiny $R$}}

\def\ii{\mbox{\tiny $i$}}
\def\l{\mbox{\tiny $l$}}

\def\P{\mbox{\tiny $P$}}

\def\j{\mbox{\tiny $j$}}
\def\mi{\mbox{\tiny $-$}}

\def\bb#1{\mbox{\footnotesize $(#1)$}}
\begin{document}

\title{Obtaining the equation of motion for a fermionic particle in a generalized Lorentz-violating system framework}% Force line breaks with \\

\author{A. E. Bernardini}
\email{alexeb@ifi.unicamp.br}
\affiliation{Instituto de F\'{\i}sica Gleb Wataghin, UNICAMP,\\
PO Box 6165, 13083-970, Campinas, SP, Brasil.}
\author{R. da Rocha}
\email{roldao.rocha@ufabc.edu.br, roldao@ifi.unicamp.br}
\affiliation{Centro de Matem\'atica, Computa\c{c}\~ao e Cogni\c{c}\~ao, Universidade Federal do ABC, 09210-170, Santo Andr\'e, SP, Brazil.}
\altaffiliation[Also at]{~Instituto de F\'{\i}sica ``Gleb Wataghin'', Unicamp, 13083-970 Campinas, SP, Brasil.}

\date{\today}% It is always \today, today,
             %  but any date may be explicitly specified

\begin{abstract}
Using a generalized procedure for obtaining the dispersion relation and the equation of motion for a propagating fermionic particle, we examine previous claims for a preferred axis at $n_{\mu}$($\equiv(1,0,0,1)$), $n^{2}=0$ embedded in the framework of very special relativity (VSR).
We show that, in a relatively high energy scale, the corresponding equation of motion is reduced to a conserving lepton number chiral equation previously predicted in the literature.
Otherwise, in a relatively low energy scale, the equation is reduced to the usual Dirac equation for a free propagating fermionic particle.
It is accomplished by the suggestive analysis of some special cases where a nonlinear modification of the action of the Lorentz group is generated by the addition of a modified conformal transformation which, meanwhile, preserves the structure of the ordinary Lorentz algebra in a very peculiar way.
Some feasible experiments, for which Lorentz violating effects here pointed out may be detectable, are suggested.
\end{abstract}

\date{\today}

\maketitle
Recently, several theoretical developments have pointed to the possibility that many empirical successes of special relativity do need not demand Lorentz invariance of the underlying theoretical framework \cite{Ame01,Mag02,Kos06,AmeDSR}.
Among the most sensitive experimental data from which such an assertion can eventually be confirmed there are the threshold anomalies in the ultra high energy cosmic ray (UHECR) protons \cite{Bie01,Tak98}, and $TeV$ photons \cite{Fin01} supposedly explained by modifications of the energy momentum dispersion relations \cite{Ame01,Mag02,Mag03}.
In this context, and also motivated by ideas from quantum gravity \cite{Smo95,Ame02}, an extension of special theory of relativity known as deformed (or doubly) special relativity (DSR) \cite{AmeDSR} was firstly proposed in manner to introduce two independent scales: the Planck scale and the velocity of light.
These theoretical advances have been paralleled by a perturbative framework developed to investigate a certain class of departures from Lorentz invariance.
For instance, Coleman and Glashow \cite{Col99,Col97} suggest spacetime translations along with exact rotational symmetry in the rest frame of the cosmic background radiation, but allow small departures from boost invariance in this frame.
In some other cases the introduction in the Lagrangian of more general Lorentz-violating terms are analyzed \cite{Kos97,Kos98,DeG06}.
Although more sensitive searches are being carried out, no decisive evidence contradicting the exact Lorentz invariance has yet been experimentally detected.
The most recent results with regard to ultra high energy protons are indeed unclear in the sense that it is uncertain whether one needs violation of Lorentz invariance for explaining the data \cite{Sch07}.

Still in the theoretical scenario, Cohen and Glashow pursue a different approach to the possible failure of Lorentz symmetry denominated {\em very special relativity} (VSR) \cite{Gla06A,Gla06B} which is based on the hypothesis that the space-time symmetry group of nature is smaller than the Poincar\'{e} group, consisting of space-time translations and one of certain subgroups of the Lorentz group.
The formalism of VSR has been expanded for studying some peculiar aspects of neutrino physics with the VSR subgroup chosen to be the 4-parameter group SIM(2) \cite{Gla06B}.
Since neutrinos are now known to be massive, several mechanisms have been contrived to remedy the absence of neutrino mass in the Standard Model Lagrangian \cite{Bil03}.
The framework of VSR admits the unconventional possibility of neutrino masses that neither violate lepton number nor require additional sterile states \cite{Gla06B}.
In the {\em Dirac} picture, lepton number is conserved with neutrinos acquiring mass via Yukawa couplings to sterile SU(2)-singlet neutrinos \cite{Zub98,Kim93}.
In the {\em Majorana} picture, lepton number is violated and neutrino masses result from a seesaw mechanism involving heavy sterile states or via dimension-6 operators resulting from unspecified new interactions \cite{Gel79,Moh86}.
In spite of not being Lorentz invariant, the lepton number conserving neutrino masses are VSR invariant.
There is no guarantee that neutrino masses have a VSR origin, but if so their sizes may be an indication of the magnitude of Lorentz-violating effects in other sectors, for instance, as a suggestion to the examination of the existence of a preferred axis in the cosmic radiation anisotropy \cite{Mag05}.

In this manuscript we intend to show how to adequate the results of VSR to a Lorentz-invariance violation system reconstructed by means of modified conformal transformations acting on the Lorentz generators.
One way to do this is to combine each boost/rotation with an specific transformation from which we introduce a preferential direction with the aid of a light-like vector defined as $n_{\mu}$($\equiv(1,0,0,1)$), $n^{2}=0$.
The transformation has to be chosen as to bring an equation of motion which recovers the dynamics of the equation introduced in \cite{Gla06B},
\begin{equation}
\left(\gamma^{\mu}p_{\mu} - \frac{m^{\2}_{\nu}}{2}\frac{\gamma^{\mu}n_{\mu}}{p_{\lambda}n^{\lambda}}\right)(1-\gamma^{\5})\nu\bb{p} = 0,
\label{pp01}
\end{equation}
which admits the unconventional possibility of neutrino masses that neither violate lepton number nor require additional sterile states.
The above dynamics requires only two degrees of freedom for a particle carrying lepton number: one for the neutrino with positive lepton number, and one for the antineutrino with negative lepton number.
The VSR neutrino at rest is necessarily an eigenstate of angular momentum in the preferred direction with eigenvalue $+1/2$.
Thus, stimulated by the ideas Cohen and Glashow \cite{Gla06A,Gla06B}, we search for convenient unitary transformations $U$ acting on the usual Lorentz generators in order to recover the equation of motion for a free propagating fermionic particle.
We expect that, in a relatively high energy scale, the corresponding equation will be reduced to the Glashow Eq.(\ref{pp01}), and, in a relatively low energy scale, it will be reduced to the usual Dirac equation for a free propagating fermionic particle.

Let us start with the definition of the momentum space $\mathcal{M}$ as the four-dimensional vector space consisting of momentum vectors $p_{\mu}$.
The ordinary Lorentz generators act as
\begin{equation}
L_{\mu\nu} = p_{\mu}\partial_{\nu} - p_{\nu}\partial_{\mu}
\label{pp02}
\end{equation}
where $\partial_{\mu} \equiv \partial/\partial p^{\mu}$, and we assume the Minkowski metric signature and that all generators are anti-Hermitian (also $\mu,\,\nu = 0,\,1,\,2,\,3$, and $i,\,j,\,k = 1,\,2,\,3$ and the velocity of the light $c = 1$).
The ordinary Lorentz algebra is constructed in terms of the usual rotations $J^{\ii}\equiv \epsilon^{\ii\j\k}L_{\j\k}$ and boosts $K^{\ii} \equiv L_{\ii\0}$ as
\begin{equation}
[J^{\ii}, K^{\j}] = \epsilon^{\ii\j\k}K_{\k};~~~~[J^{\ii}, J^{\j}] = [K^{\ii}, K^{\j}] = \epsilon^{\ii\j\k}J_{\k}
\label{pp02A}
\end{equation}
In order to introduce the nonlinear action that modifies the ordinary Lorentz generators but, however,
preserves the structure of the algebra, we suggest the following {\em ansatz} for a generalized transformation,
\begin{equation}
D \equiv (a\bb{y}\,p_{\mu} + b\bb{y}\,n_{\mu})\partial^{\mu}
\label{pp03}
\end{equation}
which acts on the momentum space as
\begin{equation}
D \circ p_{\mu} \equiv a\bb{y}\,p_{\mu} + b\bb{y}\,n_{\mu}
\label{pp04}
\end{equation}
where $y = p_{\mu} n^{\mu}$.
For such a definition, we have
\begin{equation}
\partial_{\mu} f\bb{y} = n_{\mu}\, \frac{\mbox{d}f\bb{y}}{\mbox{d}y} \equiv n_{\mu} f^{\prime}\bb{y}, ~~~~f\bb{y} = a\bb{y}, \, b\bb{y},
\label{pp05}
\end{equation}
and, consequently, $p^{\mu}\partial_{\mu} f\bb{y}= y\, f^{\prime}\bb{y}$ and $n^{\mu}\partial_{\mu} f\bb{y}= 0$.
We assume the new action can be considered to be a nonstandard and nonlinear embedding of the Lorentz group into a modified conformal group which, as we shall notice in the following for the case of main interest, despite the modifications, satisfies precisely the ordinary Lorentz algebra (\ref{pp02A}).
To exponentiate the new action, we note that
\begin{equation}
k^{\ii} = U^{^{\mi 1}}\hspace{-0.35 cm}\bb{D}\, K^{\ii}\, U\bb{D} ~~\mbox{and}
~~ j^{\ii} = U^{^{\mi 1}}\hspace{-0.35 cm}\bb{D} \,J^{\ii} \,U\bb{D}
\label{pp06}
\end{equation}
where the $y$-dependent transformation $U\bb{D}$ is given by $U\bb{D\bb{y}} \equiv{\exp[D\bb{y}]}$.
The nonlinear representation is then generated by $U\bb{D\bb{y}} \circ p_{\mu}$ and, despite not being unitary ($U\bb{D\bb{y}} \circ p_{\mu} \neq p_{\mu}$), it has to preserve the structure of the algebra.
Thus, when we assume
\begin{equation}
\left[[L_{\mu\nu},\,D\bb{y}],\,D\bb{y}\right] = 0
\label{pp07}
\end{equation}
we can reobtain a set of generators (in terms of $k_{\ii}$ and $j_{i}$) which satisfy the ordinary Lorentz algebra of (\ref{pp02A}).
At this point, in order to explicitly obtain the operator $D\bb{y}$ which satisfies the relation (\ref{pp07}), we firstly compute the commuting relation
\begin{equation}
[L_{\mu\nu},\,D\bb{y}] = \kappa_{\mu\nu}(a^{\prime}\bb{y}\,p_{\alpha} + b^{\prime}\bb{y}\,n_{\alpha})\partial^{\alpha}
+ b\bb{y} \, d_{\mu\nu},
\label{pp08}
\end{equation}
for which we have defined the parameters
\begin{eqnarray}
&&\kappa_{\mu\nu} = p_{\mu} n_{\nu} - p_{\nu} n_{\mu}\nonumber\\
&&d_{\mu\nu} = n_{\nu} \partial_{\mu} - n_{\mu} \partial_{\nu}.
\label{pp09}
\end{eqnarray}
From the above definitions we obtain the useful relations
\begin{eqnarray}
&&D \,\kappa_{\mu\nu} = a\bb{y} \kappa_{\mu\nu}\nonumber\\
&&d_{\mu\nu} \,D = a\bb{y} d_{\mu\nu}\nonumber\\
&&D \,d_{\mu\nu} = 0
\label{pp10}
\end{eqnarray}
which are essential in computing $[[L_{\mu\nu},\,D\bb{y}],\,D\bb{y}]$.
The first part of the r.h.s. of the Eq.(\ref{pp08}) then leads to the commuting relation
\begin{eqnarray}
\lefteqn{[\kappa_{\mu\nu}(a^{\prime}\bb{y}\,p_{\alpha} + b^{\prime}\bb{y}\,n_{\alpha})\partial^{\alpha},\,D\bb{y}] =}\nonumber\\
&& \kappa_{\mu\nu}\left\{\left[y (a^{^{\prime} \2}\bb{y} - a\bb{y}\, a^{\prime\prime}\bb{y}) - a\bb{y}\,a^{\prime}\bb{y}\right] p^{\alpha}\partial_{\alpha}\right.\nonumber\\
&&~~~~+ \left.\left[a^{\prime}\bb{y}(y\, b^{\prime}\bb{y} - b\bb{y}) -y\, a\bb{y}\,b^{\prime\prime}\bb{y}\right] n^{\alpha}\partial_{\alpha}\right\},
\label{pp11}
\end{eqnarray}
and the second part gives
\begin{equation}
[b\bb{y} \, d_{\mu\nu},\,D\bb{y}] = a\bb{y} (b\bb{y} - y \,b^{\prime}\bb{y}) d_{\mu\nu}.
\label{pp12}
\end{equation}
In order to satisfy the condition for preserving the Lorentz algebra (\ref{pp07}), the $y$-dependent coefficients can be obtained by evaluating the coupled ordinary differential equations:
\begin{eqnarray}
y (a^{\prime \2}\bb{y} - a\bb{y}\, a^{\prime \prime}\bb{y}) - a\bb{y}\,a^{\prime}\bb{y} = 0 &&~~~~(a.1)\nonumber\\
a^{\prime}\bb{y}(y\, b^{\prime}\bb{y} - b\bb{y}) -y\, a\bb{y}\,b^{\prime \prime}\bb{y}  = 0 &&~~~~(a.2)\nonumber\\
a\bb{y} (b\bb{y} - y \,b^{\prime}\bb{y})                                                = 0 &&~~~~(a.3)\nonumber
\end{eqnarray}
for which we have two types of solutions:

\paragraph*{Type-I}  $a\bb{y}=0$ and $\forall ~b\bb{y}$.

\paragraph*{Type-II}  $a\bb{y}= A y^{\n}, n\in \mathcal{R}$ and $b\bb{y}= B y$ where $A$ and $B$
are constants with respective dimensions given by $[[A]]\equiv m^{\mi\n}$ and $[[B]]\equiv m^{\mi \1}$.

Other choices for $U\bb{D}$($\equiv U\bb{y, p^{\2}, m^{\2}}$) are possible and lead to different boost generators,
but the proposed Type-I solution provides the simplest analytical procedure for recovering the Eq.~(\ref{pp01}).
For a Type-I solution where $a\bb{y} = 0$ and
\begin{equation}
b\bb{y} = - \frac{\alpha \,m^{\2}}{1+ 2 \alpha y},
\label{pp15}
\end{equation}
we can easily verify that
\begin{eqnarray}
\lefteqn{D\bb{y} \equiv - \frac{\alpha\,m^{\2}}{1 + 2 \alpha y}\,n_{\mu}\partial^{\mu}}\nonumber\\
&\Rightarrow&D\bb{y}\circ p_{\mu}\equiv -\frac{\alpha\,m^{\2}}{1 + 2 \alpha y}\,n_{\mu}  \nonumber\\
&\Rightarrow&U\bb{D\bb{y}}\circ p_{\mu}\equiv p_{\mu} - \frac{\alpha\,m^{\2}}{1 + 2 \alpha y}\,n_{\mu}.\nonumber\\
\label{pp16}
\end{eqnarray}
At the same time, we obtain the new generators $j_{\ii}$ and $k_{\ii}$ from Eq.~(\ref{pp06}),
\begin{eqnarray}
j_{\1}&=&J_{\1}+ 2p_{\2}\,m^{\mi\2}\,b^{^{\2}}\hspace{-0.15cm}\bb{y}\,n_{\lambda}\partial^{\lambda} + b\bb{y}\, \partial_{\2}\nonumber\\
j_{\2}&=&J_{\2}- 2p_{\1}\,m^{\mi\2}\,b^{^{\2}}\hspace{-0.15cm}\bb{y}\,n_{\lambda}\partial^{\lambda} - b\bb{y}\, \partial_{\1}\nonumber\\
j_{\3}&=&J_{\3}\nonumber\\
k_{\1}&=&K_{\1}+ 2p_{\1}\,m^{\mi\2}\,b^{^{\2}}\hspace{-0.15cm}\bb{y}\,n_{\lambda}\partial^{\lambda} + b\bb{y} \,\partial_{\1}\nonumber\\
k_{\2}&=&K_{\2}+ 2p_{\2}\,m^{\mi\2}\,b^{^{\2}}\hspace{-0.15cm}\bb{y}\,n_{\lambda}\partial^{\lambda} + b\bb{y}\, \partial_{\2}\nonumber\\
k_{\3}&=&K_{\3}+ 2(p_{\3}-p_{\0})m^{\mi\2}\, b^{^{\2}}\hspace{-0.15cm}\bb{y}\,n_{\lambda}\partial^{\lambda} + b\bb{y}\,(\partial_{\3}-\partial_{\0})
\label{pp16B}
\end{eqnarray}
from which we can reconstruct the Lorentz algebra as
\begin{equation}
[R^{\ii}, T^{\j}] = \epsilon^{\ii\j\k}T_{\k};~~~~[R^{\ii}, R^{\j}] = [T^{\ii}, T^{\j}] = \epsilon^{\ii\j\k}R_{\k}
\label{pp16C}
\end{equation}
with
\begin{eqnarray}
R_{\1(\2)}&=& \frac{1}{\sqrt{2}} (k_{\1(\2)}+\bb{-} j_{\2(\1)})\nonumber\\
T_{\1(\2)}&=& \frac{1}{\sqrt{2}} (j_{\1(\2)}+\bb{-} k_{\2(\1)})\nonumber\\
R_{\3}&=& j_{\3}\nonumber\\
T_{\3}&=& k_{\3}
\label{pp16D}
\end{eqnarray}
These transformations clearly do not preserve the usual quadratic invariant in the momentum space.
But there is a modified invariant $||U\bb{D\bb{y}}\circ p_{\mu}||^{\2} = M^{^{\2}}\hspace{-0.15cm}\bb{\alpha}$ which leads to
the following dispersion relation,
\begin{eqnarray}
||U\bb{D\bb{y}}\circ p_{\mu}||^{\2} = p^{\2} - \frac{2\, y\,m^{\2}\,\alpha}{1 + 2 \alpha y} =M^{^{\2}}\hspace{-0.15cm}\bb{\alpha}
\label{pp17}
\end{eqnarray}
Imposing the constraint $p^{\2}= m^{\2}$, which is also required by the VSR theory, we have the Casimir invariant
\begin{equation}
M^{^{\2}}\hspace{-0.15cm}\bb{\alpha} = \frac{m^{\2}}{1 + 2 \alpha y}.
\label{pp18}
\end{equation}
for which the $U$-invariance can be easily verified when we apply the transformation $U\bb{D\bb{y}}$.
The fact that the algebra of the symmetry group remains the same suggests that perhaps the standard spin connection formulation of relativity is still valid.
In this sense, the above dispersion relation can also be obtained from the dynamic equation for a fermionic particle,
\begin{eqnarray}
\lefteqn{\left[\gamma^{\mu}\left(U\bb{D\bb{y}}\circ p_{\mu}\right) - M\bb{\alpha}\right]\nu_{_{\L}}\bb{p} = 0} \nonumber\\
&&\Rightarrow\left(\gamma^{\mu}p_{\mu} - \frac{m^{\2} \alpha}{1 + 2\,\alpha\,y }\gamma^{\mu}n_{\mu} - M\bb{\alpha} \right)\nu_{_{\L}}\bb{p} = 0.
\label{pp19}
\end{eqnarray}
Alternatively, as pointed out in \cite{Mag03}, for a comparative purpose to all these classes of Lorentz-violating models,
dispersion relations may be derived from calculations in a theory such as loop quantum gravity \cite{Gam99}.

By setting $[[m]]^{\mi\1}$ values to $\alpha$, for instance, $\alpha = \pm 1/m,\,\pm m/\varepsilon^{\2}_{\P\l}$
(where $\varepsilon_{\P\l}$ is the Planck energy),
we are able to analyze the low and the high energy limits.
In the high energy limit where $\alpha y >> 1$, the Eq.~(\ref{pp19}) is reduced to
\begin{equation}
\left(\gamma^{\mu}p_{\mu} - \frac{m^{\2}}{2\,y }\gamma^{\mu}n_{\mu} - M\bb{\alpha}\right)\nu_{_{\L}}\bb{p} = 0.
\label{pp19A}
\end{equation}
and since $M^{^{\2}}\hspace{-0.15cm}\bb{\alpha}\approx \frac{m^{\2}}{2\,|\alpha\,y|} << m^{\2}$,
in spite of not being necessary, we can eliminate the dependence on $\alpha$ since
the $M\bb{\alpha}$ term becomes irrelevant in the above equation.
Thus we recover the {\em VSR Cohen-Glashow} equation (\ref{pp01})\cite{Gla06B}
and its corresponding dispersion relation, as we have proposed from the initial part of this letter.

In the low energy limit where $|\alpha y| << 1$, the Eq.~(\ref{pp19}) is reduced to
\begin{equation}
\left(\gamma^{\mu}p_{\mu} - m^{\2}\,\alpha \gamma^{\mu}n_{\mu} - m \right)\nu_{_{\L}}\bb{p} = 0,
\label{pp19B}
\end{equation}
whose the quadratic form is
\begin{equation}
\left(p^{\2} - 2 m^{\2}\,\alpha y - m^{\2}\right)\nu_{_{\L}}\bb{p} \approx  \left(p^{\2} -  m^{\2}\right)\nu_{_{\L}}\bb{p} = 0,
\label{pp19C}
\end{equation}
i.e. when $|\alpha y| << 1$ the effective contribution from the second term of Eq.~(\ref{pp19B}) is minimal and the equation can be reduced to the usual (low energy limit) {\em Dirac} equation for a free propagating particle \footnote{If we assume that fermionic particles can be classified as Dirac and Majorana particles, which depends on the way that the mass term is constructed in the Lagrangian, the equation of motion obtained in (\ref{pp19C}) describes only Majorana particles with a violating lepton number mass term ($\bb{\nu_{_{\bb{\R}}}}^{c}\bb{p} = \nu^{c}_{_{\bb{\L}}}\bb{p} =\nu_{_{\bb{\L}}}\bb{p}$).
Fundamentally, the mass term in the reduced Dirac-type equation has a mass term coming from a Lagrangian which violates the conservation of lepton number. In the VSR limit this mass term disappear.}.
Such an important result could also be reproduced in a more direct way if we initially assumed
a natural energy scale where $|m \alpha| << 1$, for instance, when $\alpha =  m/\varepsilon^{\2}_{\P\l}$.
Since $M$ is reduced to $m$, the Eq.~(\ref{pp19B}) is immediately reduced to the {\em Dirac} equation.
For all the above analysis, the dispersion relation $p^{\2}= m^{\2}$ is maintained.

By constructing $U\bb{D\bb{y}}\equiv{\exp[D\bb{y}]}$ in terms of Type-II solutions, we can find many nonlinear realizations of the action of the Lorentz group which eventually can deserve a more careful analysis.
This leads to other forms for the modified invariants and, hence, to different dispersion relations.
Just for completeness, we turn back to them in order to obtain generalized analytical expressions for $U\bb{D\bb{y}} \circ p_{\mu}$.
In particular, for Type-II solutions with $n = +1,\,-1, \, \-1/2$, we obtain
\small
\begin{eqnarray}
\lefteqn{D\bb{y} \equiv (A \, y \,p_{\mu} + B y\,n_{\mu})\partial^{\mu}}\nonumber\\
&\Rightarrow& D\bb{y}\circ p_{\mu}\equiv A \, y \,p_{\mu} + B y\,n_{\mu}  \nonumber\\
&\Rightarrow& U\bb{D\bb{y}}\circ p_{\mu}\equiv \frac{p_{\mu}}{1 - A\, y} + \frac{B \,y\, n_\mu}{1 - A\, y}\nonumber\\
&& \mbox{for}~ n = +1,~~~~~~~~~~~~~~~~~~~~~~~~~~~~~~~~~~~~~~~~~
\label{pp20A}
\end{eqnarray}
\begin{eqnarray}
\lefteqn{D\bb{y} \equiv (\frac{A}{y} p_{\mu} + B y\,n_{\mu})\partial^{\mu}}\nonumber\\
&\Rightarrow& D\bb{y}\circ p_{\mu}\equiv \frac{A}{y}p_{\mu} + B y\,n_{\mu}  \nonumber\\
&\Rightarrow& U\bb{D\bb{y}}\circ p_{\mu}\equiv p_{\mu}\left(1+ \frac{A}{y}\right) + B (A + y)n_{\mu}\nonumber\\
&& \mbox{for}~ n = -1~~~~~~~~~~~~~~~~~~~~~~~~~~~~~~~~~~~~~~~~~
\label{pp20B}
\end{eqnarray}
\begin{eqnarray}
\lefteqn{D\bb{y} \equiv (\frac{A}{\sqrt{y}}p_{\mu} + B y\,n_{\mu})\partial^{\mu}}\nonumber\\
&\Rightarrow& D\bb{y}\circ p_{\mu}\equiv \frac{A}{\sqrt{y}}p_{\mu} + B y\,n_{\mu}  \nonumber\\
&\Rightarrow& U\bb{D\bb{y}}\circ p_{\mu}\equiv \left(1+ \frac{A}{2\sqrt{y}}\right)^{^{\2}}p_{\mu} +
B \left(\sqrt{y}+ \frac{A}{2}\right)^{^{\2}}n_{\mu}\nonumber\\
&& \mbox{for}~ n = -1/2 ~~~~~~~~~~~~~~~~~~~~~~~~~~~~~~~~~~~~~~~~~
\label{pp20C}
\end{eqnarray}
\normalsize
We call upon these results in parallel works where we embed our discussion in a general formalism of conformal transformations \cite{Ber07A} and investigate the Lorentz-violating effects related to the modified dynamics introduced by them.

For the same proposal of this manuscript, we have also noticed the use of the 8-parameter subgroup ISIM(2) \cite{Gib07} with an interpretation within Finsler geometry.
In this case, translations are commutative with respect to the continuous deformations of ISIM(2), and in particular, there is an 1-parameter family of deformations denoted by DISIM$_b$(2) - a subgroup of the Weyl group.
The group DISIM$_b$(2) leaves invariant Finslerian line elements, and can be used to construct point-particle Lagrangians of Finsler form, in the context of non-linear realizations.
This approach was introduced by Ref.\cite{Bog83} and was applied for deriving a non-linear DISIM$_b$(2)-invariant generalization of the massive Dirac equation \cite{Bog04}.

To conclude, we have examined previous theoretical claims for a preferred axis at $n_{\mu}$($\equiv(1,0,0,1)$), $n^{2}=0$ in the framework of Lorentz invariance violation by generalizing the procedure for obtaining the equation of motion for a propagating fermionic particle.
We have shown that, in a relatively high energy scale, the corresponding equation of motion is reduced to a conserving lepton number chiral equation previously predicted in the literature \cite{Gla06B}, and, in a relatively low energy scale, it is reduced to the usual Dirac equation for a free propagating fermionic particle.
Effectively, one can ask to what extent the quoted solutions/equations of motion can be distinguished experimentally by data from gamma ray bursts \cite{Bie01,Tak98}, ultra high energy cosmic rays \cite{Fin01} and cosmological microwave background fluctuations \cite{Mag05}.
In particular, in order to incorporate some discrete spatial and causal structures at the Planck energy scale, the action which leads to Lorentz invariance with an invariant energy scale ($\varepsilon_{\P\l}$ or $m^{\2}/\varepsilon_{\P\l}$) can be taken into account simultaneously with the action here proposed.
Ideally, the formalism we are discussing, in the sense analogous to that of the VSR, can be used to compare experiment and theory, as well as to extrapolate between predictions of different experimental measurements.
For fermionic particles with the dynamics described by the Eq.~(\ref{pp10}), the phase space kinematic restriction is modified by a change in the relevant matrix element.
In this context, the subsequent application of our results concerns the verification of how the weak leptonic charged current $j_{\mu}$ must be modified to ensure its conservation so that we can examine the consequences for the experimental results were neutrino masses to have a purely, a dominantly or a perturbatively Lorentz-violating origin.
To summarize, we see this formalism as being part of a phenomenology of quantum gravity effects, as opposed to directly having
a fundamental significance, at the same time that, otherwise, it implies two possible phenomenologically observable modifications to neutrino physics:
(i) modifications to the predictions to neutrinoless double beta decay \cite{Ver02},
(ii) some peculiar modifications to the endpoint of the tritium beta decay \cite{Hol92,Ber07B} that is expected to be detected by
the next generation of endpoint experiments and
(iii) small modifications to the oscillation picture due to Lorentz-violating interactions that couple with the active neutrinos and eventually allow the complete explanation of neutrino data.

{\bf Acknowledgments}
The authors thank the professors C. O. Escobar and M. M. Guzzo for useful discussions and FAPESP (PD 04/13770-0 and PDJ 05/03071-0) for the financial support.

\end{document}